\documentclass[reprint, superscriptaddress,amsmath,amssymb,aps,prl]{revtex4-1}

\usepackage{graphicx}
\usepackage{xcolor}
\usepackage{physics}
\usepackage{upgreek}
\usepackage{array}
\usepackage{braket}

\usepackage{multirow}
\newcommand{\boldgreek}[1]{{\mbox{\boldmath$ {#1} $}}}
\usepackage{verbatim}
\usepackage{graphicx}
\usepackage[normalem]{ulem}
\begin{document}
\title{Sub-hertz optical transitions in excited $\mathrm{Yb}^{+}$}
\author{Patrick McMillin}
\email{pmcmi17@physics.ucla.edu}
\author{Hassan Farhat}
\author{William Liu}
\author{Wesley C. Campbell}
\affiliation{Dept.\ of Physics \& Astronomy, University of California Los Angeles, Los Angeles, CA, USA}
\affiliation{Challenge Institute for Quantum Computation, University of California Los Angeles, Los Angeles, CA, USA}
\affiliation{Center for Quantum Science and Engineering, University of California Los Angeles, Los Angeles, CA, USA}

\date{\today}

\begin{abstract}
We present the observation of three semi-forbidden transitions in singly-ionized ytterbium from the metastable $^{2}\mathrm{F}^o_{7/2}$ state. Owing to the long lifetimes of both the upper and lower states involved, these transitions are narrow and complement those already frequently used in this atom for quantum information and searches for new physics. We report the absolute frequencies of these electric quadrupole transitions, their isotope shifts, hyperfine structure, and measurements of the quadrupole transition moments. We find that the spontaneous lifetimes of these excited states are limited by slow magnetic dipole emission to lower-lying, odd-parity states, the lowest of which is accessible by laser excitation from ${}^2\mathrm{F}_{7/2}^o$ and is an attractive candidate for hosting a long-lived qubit.

\end{abstract}

\maketitle
 
The ytterbium atom is a popular platform for precision measurement 
\cite{tofful_171yb_2024, stenger_absolute_2001, roberts_measurement_1999, Barber_LS} and quantum information science \cite{moses_race-track_2023,Muniz_Universal_YBGates} due mostly to its combination of ease of control and desirable nuclear properties.  Both the neutral and singly-ionized charge states have multiple metastable states that furnish them with optical transitions with sub-$10\,\mathrm{Hz}$ natural linewidths, which are used for optical clocks \cite{Hoyt_Yb_Clock, Huntemann_Yb+_clock} and precision tests of physics beyond the Standard Model (BSM)\cite{Ono_Yb_Kingplot,sanner_CloclLor,dreissen_lorentz}.

In particular, the presence of multiple semi-forbidden transitions and an abundance of stable isotopes for the ytterbium nuclide has allowed isotope-shift determinations with high accuracy. A recent study found a statistically significant King-plot nonlinearity in a search for signatures of BSM physics \cite{hur_evidence_2022}. The effects of higher-order Standard Model terms may be removed by adding additional narrow transitions and isotope pairs to King-plot analysis, which is critical for isolating the presence of BSM physics through isotope-shift measurements. Further, additional optically narrow transitions can be used as alternative optical clocks and qubit transitions.  

The extremely long lifetime ($\approx2$ years, \cite{R_Lange_lifetime}) of the metastable $^{2}\mathrm{F}^o_{7/2}$ state in $\mathrm{Yb}^+$ has been identified as a unique feature that can serve as a second ground state of the ion for quantum information processing. A qubit defined on the zero-field hyperfine clock states ($M_F=0$) in the $^{2}\mathrm{F}_{7/2}^o$ state of $^{171}\mathrm{Yb}^{+}$ has been shown to have excellent state preparation and measurement (SPAM) fidelity \cite{roman_expanding_2021, yang_realizing_2022}, and could be implemented as part of the ``$omg$ protocol'' for trapped ion quantum information processing \cite{OMG}. However, a key ingredient for implementing this is the development of qubit-motion coupling of the $m$-type qubit in the $^{2}\mathrm{F}_{7/2}^o$ state, which can be achieved via a narrow optical transition to a metastable state. The spectrum of Yb$^{+}$ contains dozens of excited odd-parity states arising from the low-lying multi-electron configuration of $4f^{13}(^{2}\mathrm{F}^{o}_{7/2})5d6s$, generating multiple metastable states described by the $J_{1}K$-coupling scheme.

In this letter, we demonstrate the presence of additional sub-hertz optical transitions between metastable levels of $\mathrm{Yb}^+$. We report laser spectroscopy of electric-quadrupole ($E2$) transitions from ${}^2\mathrm{F}_{7/2}^o$ to three $J_1 K$-coupled states at red and near-infrared wavelengths. Isotope-shift measurements for these transitions are provided, as well as the ${}^{171}\mathrm{Yb}^+$ hyperfine $A$ coefficients of the excited levels. We report the ${}^2\mathrm{F}_{7/2}^o \leftrightarrow {}^3[7/2]_{9/2}^o$ transition moment and radiative lifetime of the excited state showing a sub-hertz optical transition linewidth. Further, we report the upper limits on the $E2$ Einstein $A$ coefficients of the ${}^2\mathrm{F}_{7/2}^o \leftrightarrow {}^3[9/2]_{9/2}^o$ and the ${}^2\mathrm{F}_{7/2}^o \leftrightarrow {}^1[11/2]_{11/2}^o$ transitions. These results, which demonstrate transitions with linewidths narrower than those involving the ${}^2\mathrm{D}_J$ states in $\mathrm{Yb}^+$, suggest that metastable $\mathrm{Yb}^+$ may prove useful for applications in clockwork, precision measurement, and quantum information processing. Due to the small $E2$ transition moments, we have found that the spontaneous emission lifetime of the excited states investigated here are limited by magnetic dipole ($M1$) decays to other states with the same electron configuration. 

The selected $J_{1}K$-coupled excited states in this work have $J>7/2$, odd parity, and the same $4f^{13}$ core-electron configuration as $^{2}\mathrm{F}^o_{7/2}$.  Similar to the ${}^{2}\mathrm{S}_{1/2} \leftrightarrow {}^2\mathrm{D}_J$ transitions, each $E2$ transition from the $^{2}\mathrm{F}^o_{7/2}$ state to a $J_{1}K$-coupled excited state involves a single-electron excitation from the $6s$ to $5d$ subshells. However, unlike transitions originating from the ${}^{2}\mathrm{S}_{1/2}$ state, these transitions occur between states containing a spinful hole in the $4f$ subshell, and can be spin-forbidden for the outer electron pair. 

For the spectroscopic identification of these new transitions, single ytterbium ions were trapped in a radio frequency Paul trap with a magnetic field of $4.00(7)\text{ }\mathrm{G}$ applied to define the quantization axis and destabilize coherent dark states. To initialize the ion in the ${}^2\mathrm{F}_{7/2}^o$ manifold, laser light is applied to transfer population from the $^{2}\mathrm{S}_{1/2}$ state to the $^{2}\mathrm{D}_{5/2}$ state ($\tau=7.2\,\mathrm{ms}$), which favorably leaks to the $^{2}\mathrm{F}^o_{7/2}$ state with a branching fraction of $0.83(3)$ \cite{tan_precision_2021}. The linear polarization and $k$ vector of this light are both set perpendicular to the magnetic field ($\mathbf{k}\perp\mathbf{B}$ and  $\hat{\boldsymbol{\epsilon}}\perp\mathbf{B}$) to select for $\Delta m_{J} = \pm2$ transitions \cite{CampbellMultipoles}. Excited state readout is facilitated by a laser tone driving the $^{2}\mathrm{F}^o_{7/2} \leftrightarrow \, ^{1}[3/2]^o_{3/2}$ $E2$ transition, which causes population to preferentially decay back to the $^{2}\mathrm{S}_{1/2}$ ground state. The $^{2}\mathrm{F}^o_{7/2}$ population is measured by observing the presence or lack of laser-induced fluorescence on the strong $^{2}\mathrm{S}_{1/2} \leftrightarrow \, ^{2}\mathrm{P}^{o}_{1/2}$ transition. 

\begin{figure}
    \includegraphics[scale=1]{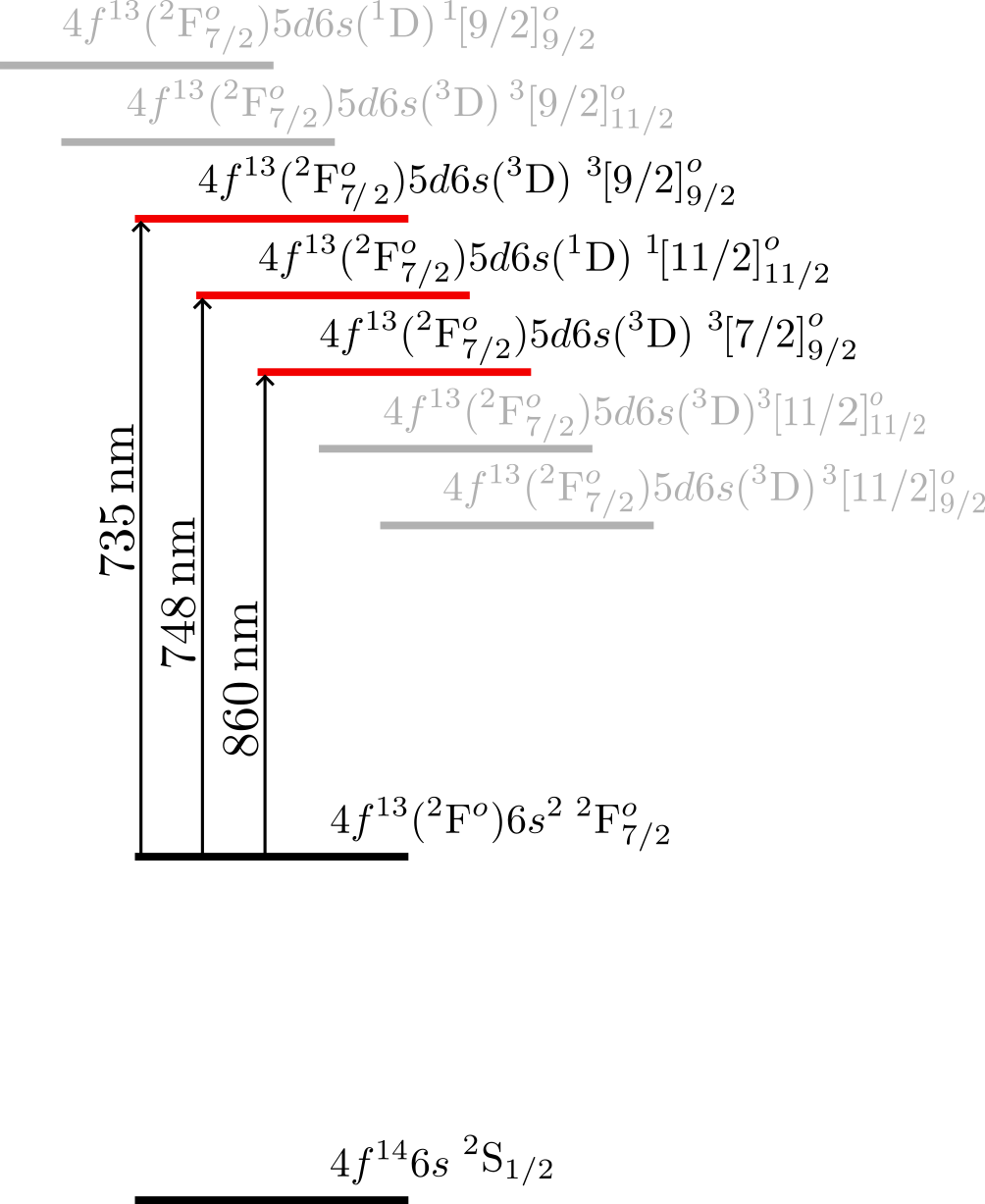}
    \caption{Level diagram displaying driven transitions from the $^{2}\mathrm{F}^o_{7/2}$ state in $\mathrm{Yb^+}$.}
    \label{fig:GROT}
\end{figure}

We implement a pulsed spectroscopy sequence with the spectroscopy laser's frequency controlled by a double-passed acousto-optic modulator to determine the frequency of the transitions. Ion population is first prepared in the $^{2}\mathrm{F}^o_{7/2}$ state by exciting the ${}^2\mathrm{S}_{1/2} \leftrightarrow {}^2\mathrm{D}_{5/2}$ transition for 200 ms. A 5 ms spectroscopy pulse at variable frequency then drives population from $^{2}\mathrm{F}^o_{7/2}$ to an excited $J_{1}K$-coupled state. Spectroscopy light propagates with $\mathbf{k}\perp\mathbf{B}$ and linear polarization $\hat{\boldsymbol{\epsilon}}\parallel\mathbf{B}$, selecting for $\Delta m_J=\pm1$ transitions \cite{CampbellMultipoles}. We then immediately depopulate the $^{2}\mathrm{F}^{o}_{7/2}$ state by driving the $^{2}\mathrm{F}^{o}_{7/2} \leftrightarrow \, ^{1}[3/2]^{o}_{3/2}$ transition for 50 ms, and then perform state detection with 1 ms of light applied on the $^{2}\mathrm{S}_{1/2} \leftrightarrow \, ^{2}\mathrm{P}^{o}_{1/2}$ transition. Ion fluorescence indicates failed transfer out of the $^{2}\mathrm{F}^o_{7/2}$ manifold, while the absence of ion fluorescence indicates successful transfer.

\begin{figure*}
    \includegraphics[scale=0.44]{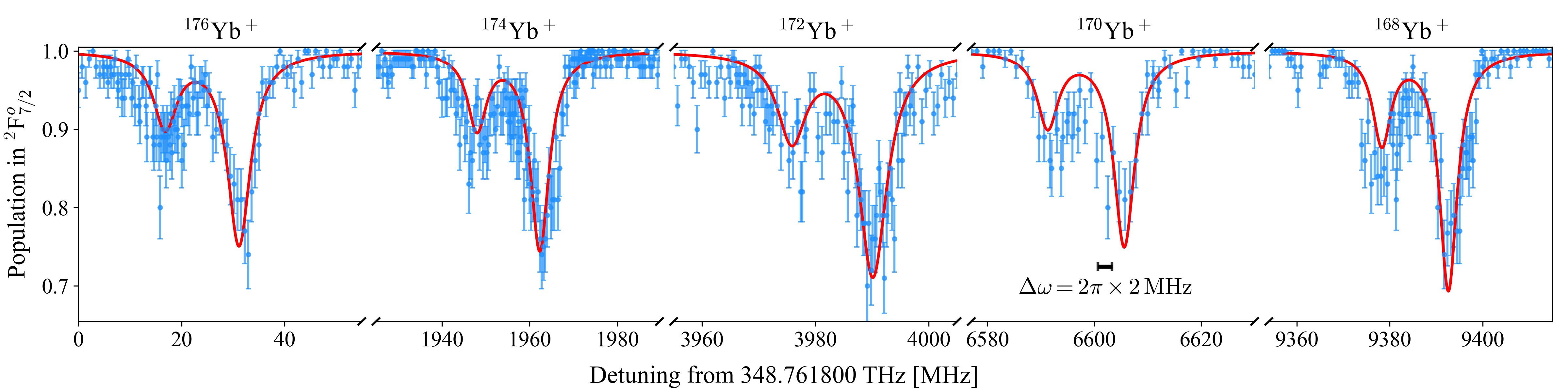}
    \caption{$^2\mathrm{F}_{7/2}^o \leftrightarrow {}^3[7/2]_{9/2}^o$, $|\Delta m_J|=1$ transition spectra across all stable even isotopes of Yb$^+$. The population distribution across the Zeeman sublevels of the $^2\mathrm{F}_{7/2}^o$ state results in the asymmetry between the $\Delta m_J=-1$ (lower frequency) feature and $\Delta m_J=+1$ (higher frequency) feature in each isotope. The linewidth of our spectroscopy laser is represented by the scale bar in the ${}^{170}\mathrm{Yb}^+$ subplot.}
    \label{fig:Spectrum}
\end{figure*}

\begin{table*}
    \centering
    \begin{tabular}{|c|c|c|c|}
    \hline
        Isotope  & $^{3}[7/2]^o_{9/2}$ (THz) & $^{1}[11/2]^o_{11/2}$  (THz) & $^{3}[9/2]^o_{9/2}$  (THz)\\
        \hline \hline
        168 & 348.77119(3) & Not Measured & Not Measured \\ \hline
        170 & 348.76841(3) & 400.71232(3) & 407.73232(3) \\ \hline
        172 & 348.76579(3) & 400.70937(3) & 407.72974(3) \\ \hline
        174 & 348.76376(3) & 400.70708(3) & 407.72774(3) \\ \hline 
        176 & 348.76183(3) & 400.70490(3) & 407.72583(3) \\
        \hline
    \end{tabular}
    \caption{Absolute frequency measurements of the newly observed transitions.  
    }
    \label{tab:placeholder}
\end{table*}

Recording the number of failed and successful events as a function of laser frequency generates the population remaining in the $^{2}\mathrm{F}^o_{7/2}$ manifold. Detailed linescans for the $^{2}\mathrm{F}^{o}_{7/2} \leftrightarrow \, ^{3}[7/2]^{o}_{9/2}$ transition in each stable even isotope of $\mathrm{Yb}^+$ are shown in figure \ref{fig:Spectrum}. The uncertainties of the transition frequencies are on the order of 10s of MHz primarily due to the incoherent population preparation in the Zeeman sublevels of $^{2}\mathrm{F}^{o}_{7/2}$ state and the laser linewidth ($\approx 2\, \mathrm{MHz}$). 

\begin{figure}
    \includegraphics[scale=0.5]{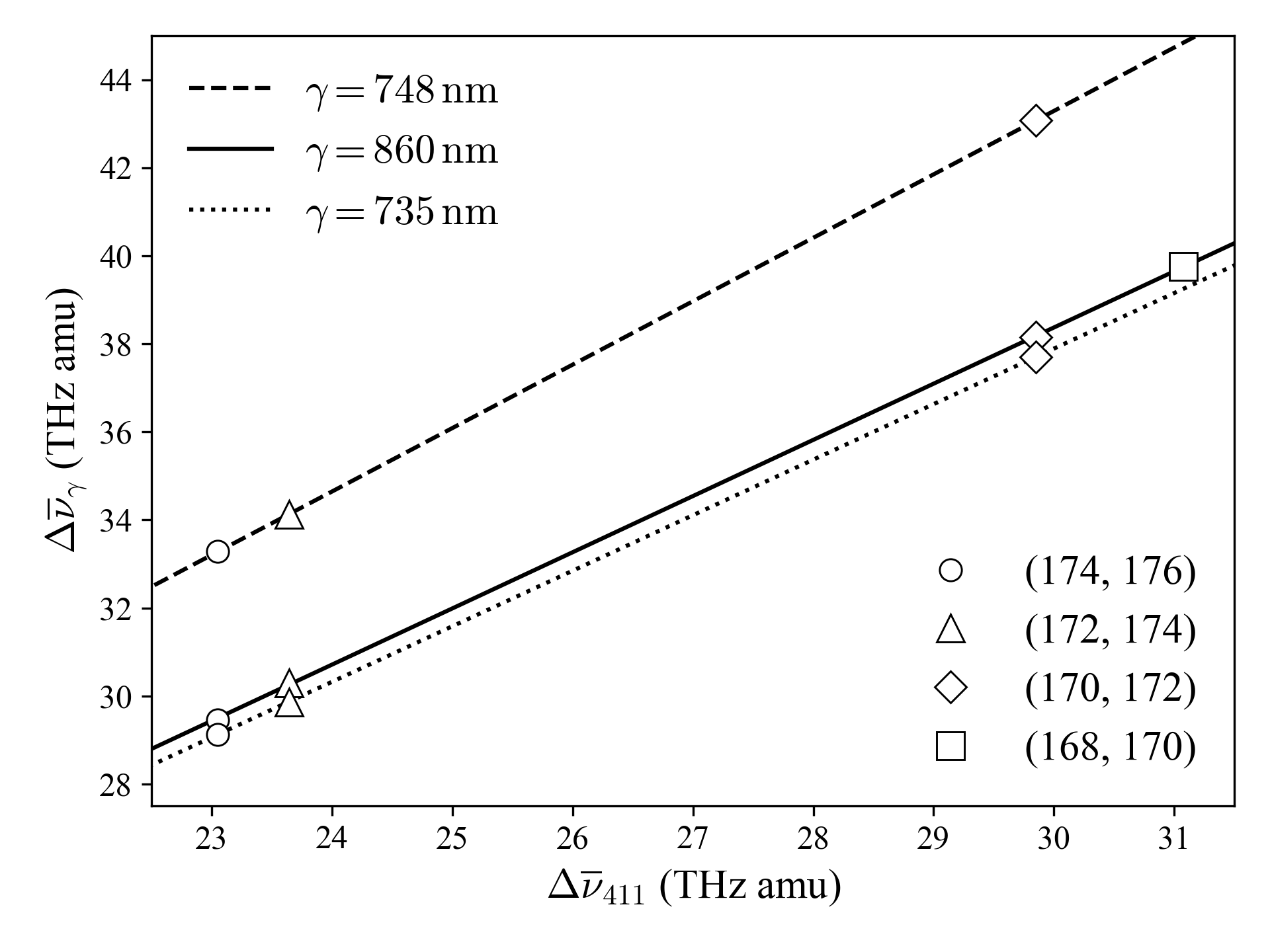}
    \caption{King plot comparing the modified isotope shifts of the driven transitions against those of the $^{2}\mathrm{S}_{1/2} \leftrightarrow \, ^{2}\mathrm{D}_{5/2}$ measured by the Vuletić group \cite{Hur_thesis}. Note that only the $^{2}\mathrm{F}^{o}_{7/2} \leftrightarrow \, ^{3}[7/2]^{o}_{9/2}$ transition was measured for $^{168} \mathrm{Yb}^+.$}
    \label{fig:KingPlot} 
\end{figure}

A King plot for the  $^{2}\mathrm{F}^{o}_{7/2} \leftrightarrow \, ^{3}[7/2]^{o}_{9/2}$, $^{2}\mathrm{F}^{o}_{7/2} \leftrightarrow \, ^{1}[11/2]^{o}_{11/2}$, and $^{2}\mathrm{F}^{o}_{7/2} \leftrightarrow \, ^{3}[9/2]^{o}_{9/2}$ transitions compared to the $^{2}\mathrm{S}_{1/2} \leftrightarrow \, ^{2}\mathrm{D}_{5/2}$ transition for even isotopes of $\mathrm{Yb^+}$ are presented in figure \ref{fig:KingPlot}. For $^{168}\mathrm{Yb}^{+}$, only the $^{2}\mathrm{F}^{o}_{7/2} \leftrightarrow \, ^{3}[7/2]^{o}_{9/2}$ transition was measured.

\begin{figure}
    \includegraphics[scale=1.3]{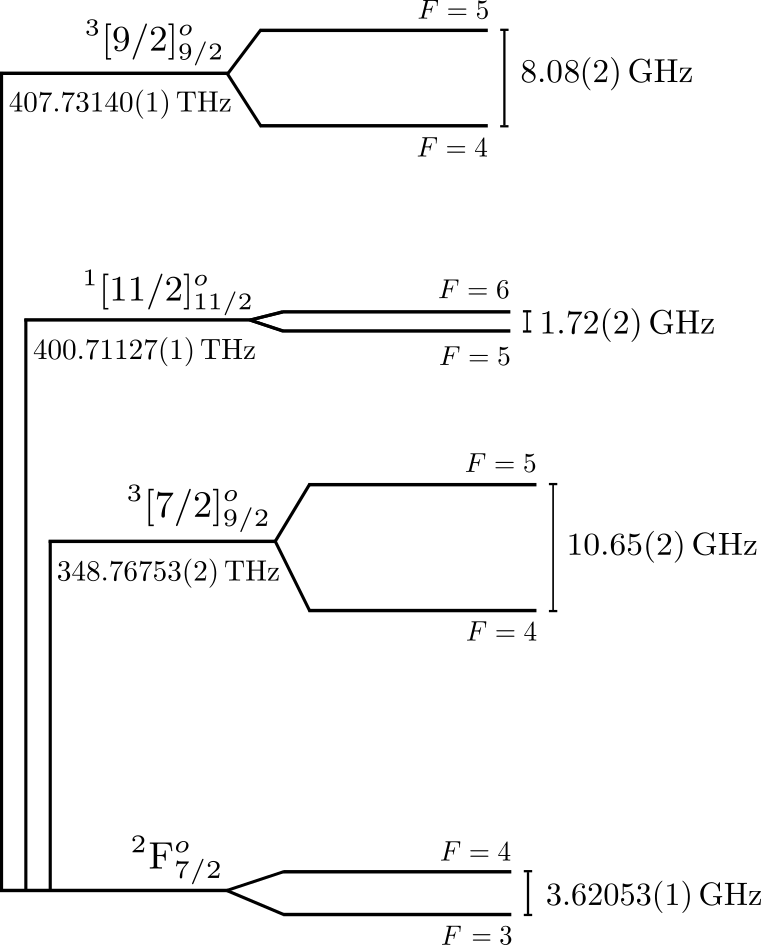}
    \caption{Hyperfine structure of $^{171}\mathrm{Yb}^+$ of the excited state and the $^{2}\mathrm{F}_{7/2}^o$. The centroid frequency of each transition is presented along with the hyperfine splitting of each excited state.}
    \label{fig:centriods}
\end{figure}

In order to perform hyperfine spectroscopy of the $J_{1}K$-coupled states in $^{171}\mathrm{Yb}^{+}$, we leverage hyperfine selection rules. We simplify the determination of the sign of the excited states' hyperfine coefficients by initially populating only one hyperfine manifold ($F$=3) in the $^{2}\mathrm{F}_{7/2}^o$ state via the $^{2}\mathrm{S}_{1/2}, F=0 \rightarrow \, ^{2}\mathrm{D}_{5/2}, F=2$ transition. We measured the hyperfine $A$ coefficients of the $^{3}[7/2]^{o}_{9/2}$, $^{1}[11/2]^{o}_{11/2}$, and $^{3}[9/2]^{o}_{9/2}$ states to be 2131(5), 288(3), and 1616(4) MHz, respectively. Hyperfine splittings were then verified by populating the $F=4$ manifold in the $^{2}\mathrm{F}^o_{7/2}$ state and driving the available transitions. Figure \ref{fig:centriods} displays the centroid frequencies of the transitions, along with the hyperfine splittings of the excited states.

The odd-parity excited states investigated here can decay via magnetic dipole ($M1$) transitions to other odd-parity states, potentially causing the population to become stuck in inaccessible states. Due to the imperfect description of these states by the $J_{1}K$ coupling scheme, second-leading-order contributions to the terms must be considered when evaluating the $M1$ decay pathways. The $M1$ transition strengths can be calculated by expressing the magnetic dipole operator $\boldgreek{\mu}$ as the sum of operators $\boldgreek{\mu}_\mathrm{c}$ and $\boldgreek{\mu}_{\mathrm{o}}$ that act on the core and outer electrons, respectively. The calculated $M1$ lifetimes of the $^{3}[7/2]^o_{9/2}$, $^{1}[11/2]^o_{11/2}$, and $^{3}[9/2]^o_{9/2}$ states are $9.9$, $0.41$, and $0.24$ seconds respectively. Many $M1$ decay products are themselves metastable, as an $M1$ decay event frequently leads to ``dark" ions, which have taken up to 45 minutes to re-enter the Doppler cooling cycle. The shorter $M1$ lifetimes of the $^{1}[11/2]^{o}_{11/2}$ and $^{3}[9/2]^{o}_{9/2}$ states, along with the long $M1$ lifetimes of their decay products, hampered our ability to measure a finite $E2$ moment for these transitions.  

We measured the $E2$ Einstein $A$ coefficient of the $^{2}\mathrm{F}^{o}_{7/2} \leftrightarrow \, ^{3}[7/2]^{o}_{9/2}$ transition using our laser beam parameters and the measured transfer rate. Transfer rate experiments are identical to frequency determination experiments in their sequence structure. However, the spectroscopy laser frequency remains fixed while the pulse time is varied. The remaining population follows simple rate equations, as the spectroscopy laser linewidth is much larger than the transition linewidth.

\begin{figure}
    \includegraphics[scale=0.79]{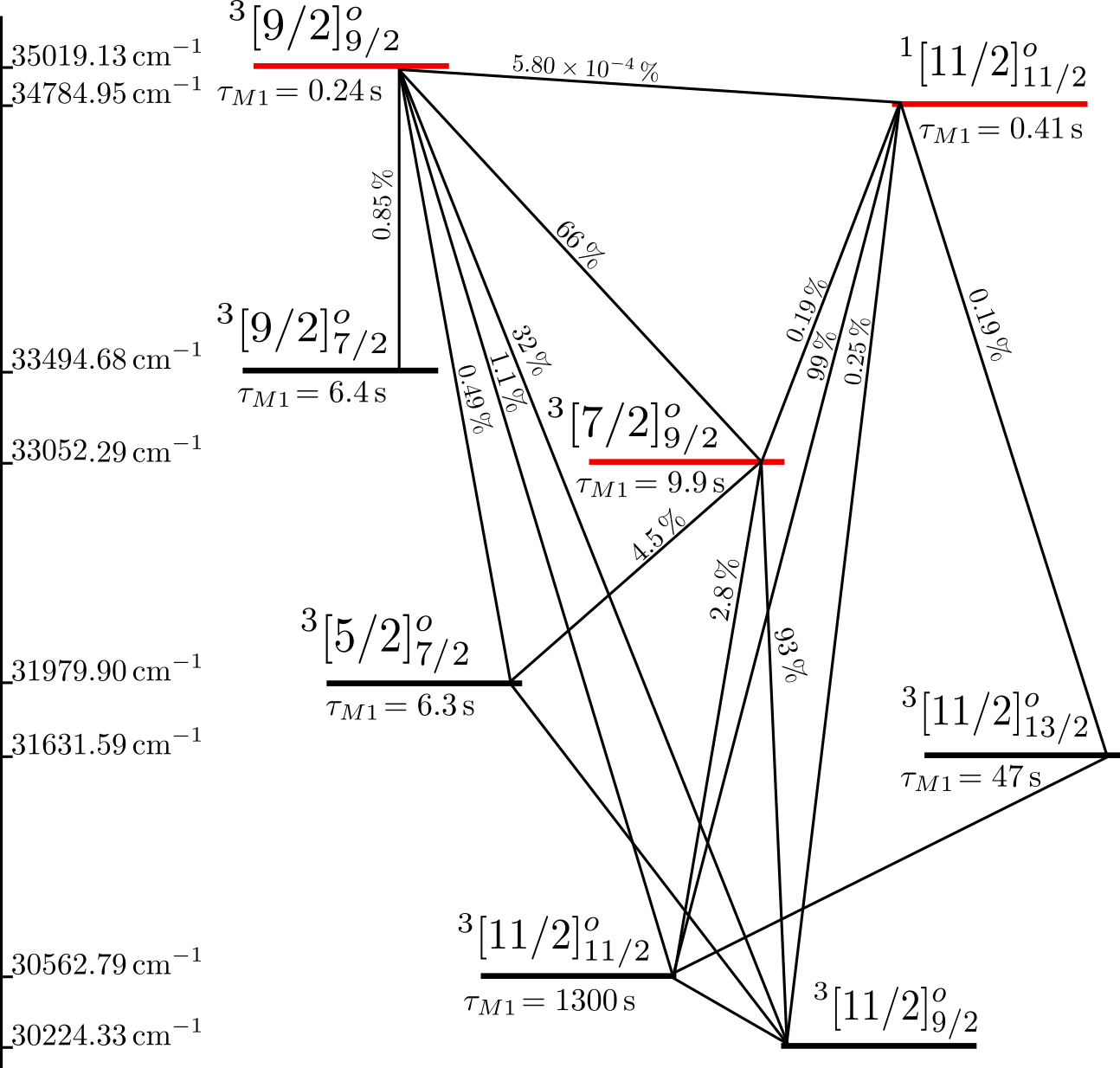}
    \caption{$M1$ decay pathways from ${}^3[7/2]^o_{9/2}$, ${}^1[11/2]^o_{11/2}$, and ${}^3[9/2]^o_{9/2}$. The presented $M1$ lifetimes and branching ratios are theoretically calculated. The energy levels are sourced from \cite{NIST_ASD}.
    }
    \label{fig:decays}
\end{figure}

\begin{figure}
    \includegraphics[scale=0.33]{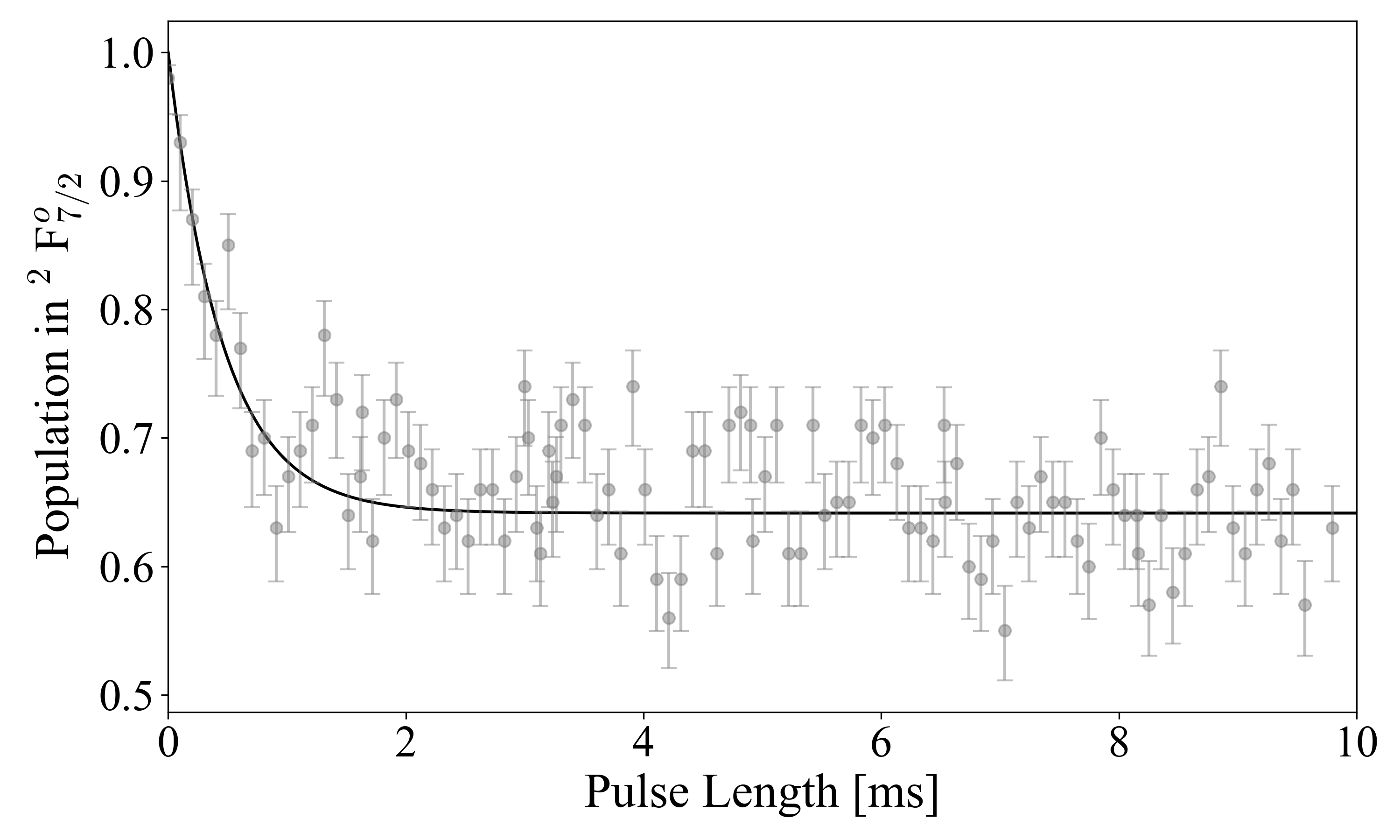}
    \caption{Population remaining in the $^{2}\mathrm{F}^o_{7/2}$ state as a function of spectroscopy laser illumination time. The data are fit to simple rate equations, allowing extraction of the transition moment when combined with our laser parameters.
    }
    \label{fig:TransferRate}
\end{figure}

Results are fitted to a two-level system of rate equations representing the population in the $^{2}\mathrm{F}^o_{7/2}$ state ($N_\mathrm{F}$) and population in the excited $^{3}[7/2]^{o}_{9/2}$ state ($N_\mathrm{B}$),
\begin{align*}
    \dot{N}_\mathrm{F}(t) &= A_{E2} N_\mathrm{B}(t)+ B\rho(\omega) (N_\mathrm{B}(t) -f N_\mathrm{F}(t))\\ 
    \dot{N}_\mathrm{B}(t) &= -\dot{N}_\mathrm{F}(t)
\end{align*}
where $\rho (\omega)$ is the energy spectral density of the applied laser light, and $f$ is the fraction of population addressed by the laser. By applying this model to the $^{2}\mathrm{F}^{o}_{7/2} \leftrightarrow \, ^{3}[7/2]^{o}_{9/2}$ transition, we measure the $E2$ Einstein $A$ coefficient to be $A_{E2} = 6(3)\times 10^{-3} \text{ }\mathrm{s^{-1}}$ and the corresponding reduced matrix element (reduced to $J$) to be $1.6(9)\text{ } ea_0^2$. Using this reduced matrix element as an upper bound for the  $^{2}\mathrm{F}^{o}_{7/2} \leftrightarrow \, ^{3}[9/2]^{o}_{9/2}$ transition, we find an estimate for the $E2$ Einstein $A$ coefficient to be $\approx 10\times 10^{-3} \text{ }\mathrm{s^{-1}}$. For the $^{2}\mathrm{F}^{o}_{7/2} \leftrightarrow \, ^{1}[11/2]_{11/2}$ transition, we estimate an $E2$ Einstein $A$ coefficient of $\approx 1$ s$^{-1}$. The reported $E2$ transition rates show that the radiative lifetimes of the excited states are limited by $M1$ decays to other $J_{1}K$-coupled states. The measured transition matrix element is an order of magnitude smaller than that of the $^{2}\mathrm{S}_{1/2} \leftrightarrow \, ^{2}\mathrm{D}_{3/2}$ and $^{2}\mathrm{S}_{1/2} \leftrightarrow \, ^{2}\mathrm{D}_{5/2}$ transitions (10.1$\text{ } ea_0^2$ and 12.5$\text{ } ea_0^2$, respectively \cite{shao_precision_2023, tan_precision_2021}). Due to the change in electron configuration in these transitions, we expect the $M1$ coupling between the $^{2}\mathrm{F}^{o}_{7/2}$ and excited $J_{1}K$-coupled states to be much smaller than the contributions of the $E2$ coupling.

The main limitation in the presented measurements is the lack of single Zeeman state preparation in the $^{2}\mathrm{F}^o_{7/2}$ state, which can be overcome in the spinless isotopes by applying an additional laser tone on the $^{2}\mathrm{F}^{o}_{7/2} \leftrightarrow \, ^{1}[3/2]^{o}_{3/2}$ transition. If the light is applied during state preparation of the $^{2}\mathrm{F}^o_{7/2}$ state such that $\Delta m_{J}=\pm 2$ transitions are not allowed (i.e. $\mathbf{k}\perp\mathbf{B}$, $\boldgreek{\hat{\epsilon}}\parallel\mathbf{B}$), population will be stranded in the $m_{J}=\pm 7/2$ sublevels in the $^{2}\mathrm{F}^o_{7/2}$ state. In $^{171}\mathrm{Yb}^{+}$, heralded state preparation can be performed by applying resonant microwaves in conjunction with highly manifold-selective laser tones to prepare the $m_{F}=0$ state with high fidelity \cite{roman_expanding_2021}. For preparing the $m$-type qubit in $^{171}\mathrm{Yb}^{+}$, this step is already necessary when populating the $^{2}\mathrm{F}^o_{7/2}$ incoherently. \par

Frequent decays from the excited $J_{1}K$-coupled states may necessitate the use of additional repump tones for some applications to ensure that optical cycling can be recovered after an $M1$ decay event. Utilizing a metastable state to perform qubit-motion coupling may be more attractive than a two-photon, $E1$-mediated transition due to the high number of spontaneous emission channels available to the excited even-parity $(J_1, J_2)$-coupled states in Yb$^{+}$. The $M1$ decay pathways from the higher-lying $J_{1}K$ states may prove to have large qubit-subspace leakage errors. Therefore, the $^{2}\mathrm{F}^{o}_{7/2} \leftrightarrow \, ^{3}[11/2]^{o}_{11/2}$ and $^{2}\mathrm{F}^{o}_{7/2} \leftrightarrow \, ^{3}[11/2]^{o}_{9/2}$ transitions are attractive candidates for implementing qubit-motion coupling owing to the readily available laser sources and the extremely long $M1$ lifetimes of these two low-lying excited states, which are depicted as the two lowest levels in figure \ref{fig:decays}.  

In order to further reduce the systematics caused by higher-order SM terms in isotope shift spectroscopy, more transitions must be added to the King-plot analysis in conjunction with additional isotope pairs. Yb has five stable spinless isotopes, but also has many long lived spinless radioactive isotopes. In particular $^{164}$Yb, $^{166}$Yb, and $^{178}$Yb have half-lives of $76 \, \mathrm{mins}$, $57 \, \mathrm{hrs}$, and $74 \, \mathrm{mins}$ respectively \cite{Kondev_2021}, which are sufficient for laser spectroscopy. The fact that single ions can be isolated and trapped means that atomic sources for these radioactive isotopes can remain at low activity \cite{Hucul_spectroscopy_2017}, and be efficiently utilized \cite{Fan_laser_2023}. 

\begin{acknowledgments}
The authors acknowledge Michael K\"ohl and Vladan Vuletić for helpful discussions.  This work was supported by ARO W911NF-24-S-0004, NSF PHY-2207985 and OMA-2016245, and the Gordon and Betty Moore Foundation DOI: 10.37807/GBMF11566.
\end{acknowledgments}

\bibliography{main}

\begin{thebibliography}{24}%
\makeatletter
\providecommand \@ifxundefined [1]{%
 \@ifx{#1\undefined}
}%
\providecommand \@ifnum [1]{%
 \ifnum #1\expandafter \@firstoftwo
 \else \expandafter \@secondoftwo
 \fi
}%
\providecommand \@ifx [1]{%
 \ifx #1\expandafter \@firstoftwo
 \else \expandafter \@secondoftwo
 \fi
}%
\providecommand \natexlab [1]{#1}%
\providecommand \enquote  [1]{``#1''}%
\providecommand \bibnamefont  [1]{#1}%
\providecommand \bibfnamefont [1]{#1}%
\providecommand \citenamefont [1]{#1}%
\providecommand \href@noop [0]{\@secondoftwo}%
\providecommand \href [0]{\begingroup \@sanitize@url \@href}%
\providecommand \@href[1]{\@@startlink{#1}\@@href}%
\providecommand \@@href[1]{\endgroup#1\@@endlink}%
\providecommand \@sanitize@url [0]{\catcode `\\12\catcode `\$12\catcode `\&12\catcode `\#12\catcode `\^12\catcode `\_12\catcode `\%12\relax}%
\providecommand \@@startlink[1]{}%
\providecommand \@@endlink[0]{}%
\providecommand \url  [0]{\begingroup\@sanitize@url \@url }%
\providecommand \@url [1]{\endgroup\@href {#1}{\urlprefix }}%
\providecommand \urlprefix  [0]{URL }%
\providecommand \Eprint [0]{\href }%
\providecommand \doibase [0]{http://dx.doi.org/}%
\providecommand \selectlanguage [0]{\@gobble}%
\providecommand \bibinfo  [0]{\@secondoftwo}%
\providecommand \bibfield  [0]{\@secondoftwo}%
\providecommand \translation [1]{[#1]}%
\providecommand \BibitemOpen [0]{}%
\providecommand \bibitemStop [0]{}%
\providecommand \bibitemNoStop [0]{.\EOS\space}%
\providecommand \EOS [0]{\spacefactor3000\relax}%
\providecommand \BibitemShut  [1]{\csname bibitem#1\endcsname}%
\let\auto@bib@innerbib\@empty
\bibitem [{\citenamefont {Tofful}\ \emph {et~al.}(2024)\citenamefont {Tofful}, \citenamefont {Baynham}, \citenamefont {Curtis}, \citenamefont {Parsons}, \citenamefont {Robertson}, \citenamefont {Schioppo}, \citenamefont {Tunesi}, \citenamefont {Margolis}, \citenamefont {Hendricks}, \citenamefont {Whale}, \citenamefont {Thompson},\ and\ \citenamefont {Godun}}]{tofful_171yb_2024}%
  \BibitemOpen
  \bibfield  {author} {\bibinfo {author} {\bibfnamefont {A.}~\bibnamefont {Tofful}}, \bibinfo {author} {\bibfnamefont {C.~F.~A.}\ \bibnamefont {Baynham}}, \bibinfo {author} {\bibfnamefont {E.~A.}\ \bibnamefont {Curtis}}, \bibinfo {author} {\bibfnamefont {A.~O.}\ \bibnamefont {Parsons}}, \bibinfo {author} {\bibfnamefont {B.~I.}\ \bibnamefont {Robertson}}, \bibinfo {author} {\bibfnamefont {M.}~\bibnamefont {Schioppo}}, \bibinfo {author} {\bibfnamefont {J.}~\bibnamefont {Tunesi}}, \bibinfo {author} {\bibfnamefont {H.~S.}\ \bibnamefont {Margolis}}, \bibinfo {author} {\bibfnamefont {R.~J.}\ \bibnamefont {Hendricks}}, \bibinfo {author} {\bibfnamefont {J.}~\bibnamefont {Whale}}, \bibinfo {author} {\bibfnamefont {R.~C.}\ \bibnamefont {Thompson}}, \ and\ \bibinfo {author} {\bibfnamefont {R.~M.}\ \bibnamefont {Godun}},\ }\href {\doibase 10.1088/1681-7575/ad53cd} {\bibfield  {journal} {\bibinfo  {journal} {Metrologia}\ }\textbf {\bibinfo {volume} {61}},\ \bibinfo {pages} {045001} (\bibinfo {year} {2024})}\BibitemShut
  {NoStop}%
\bibitem [{\citenamefont {Stenger}\ \emph {et~al.}(2001)\citenamefont {Stenger}, \citenamefont {Tamm}, \citenamefont {Haverkamp}, \citenamefont {Weyers},\ and\ \citenamefont {Telle}}]{stenger_absolute_2001}%
  \BibitemOpen
  \bibfield  {author} {\bibinfo {author} {\bibfnamefont {J.}~\bibnamefont {Stenger}}, \bibinfo {author} {\bibfnamefont {C.}~\bibnamefont {Tamm}}, \bibinfo {author} {\bibfnamefont {N.}~\bibnamefont {Haverkamp}}, \bibinfo {author} {\bibfnamefont {S.}~\bibnamefont {Weyers}}, \ and\ \bibinfo {author} {\bibfnamefont {H.~R.}\ \bibnamefont {Telle}},\ }\href {\doibase 10.1364/OL.26.001589} {\bibfield  {journal} {\bibinfo  {journal} {Optics Letters}\ }\textbf {\bibinfo {volume} {26}},\ \bibinfo {pages} {1589} (\bibinfo {year} {2001})}\BibitemShut {NoStop}%
\bibitem [{\citenamefont {Roberts}\ \emph {et~al.}(1999)\citenamefont {Roberts}, \citenamefont {Taylor}, \citenamefont {Gateva-Kostova}, \citenamefont {Clarke}, \citenamefont {Rowley},\ and\ \citenamefont {Gill}}]{roberts_measurement_1999}%
  \BibitemOpen
  \bibfield  {author} {\bibinfo {author} {\bibfnamefont {M.}~\bibnamefont {Roberts}}, \bibinfo {author} {\bibfnamefont {P.}~\bibnamefont {Taylor}}, \bibinfo {author} {\bibfnamefont {S.~V.}\ \bibnamefont {Gateva-Kostova}}, \bibinfo {author} {\bibfnamefont {R.~B.~M.}\ \bibnamefont {Clarke}}, \bibinfo {author} {\bibfnamefont {W.~R.~C.}\ \bibnamefont {Rowley}}, \ and\ \bibinfo {author} {\bibfnamefont {P.}~\bibnamefont {Gill}},\ }\href {\doibase 10.1103/PhysRevA.60.2867} {\bibfield  {journal} {\bibinfo  {journal} {Physical Review A}\ }\textbf {\bibinfo {volume} {60}},\ \bibinfo {pages} {2867} (\bibinfo {year} {1999})}\BibitemShut {NoStop}%
\bibitem [{\citenamefont {Barber}\ \emph {et~al.}(2008)\citenamefont {Barber}, \citenamefont {Stalnaker}, \citenamefont {Lemke}, \citenamefont {Poli}, \citenamefont {Oates}, \citenamefont {Fortier}, \citenamefont {Diddams}, \citenamefont {Hollberg}, \citenamefont {Hoyt}, \citenamefont {Taichenachev},\ and\ \citenamefont {Yudin}}]{Barber_LS}%
  \BibitemOpen
  \bibfield  {author} {\bibinfo {author} {\bibfnamefont {Z.~W.}\ \bibnamefont {Barber}}, \bibinfo {author} {\bibfnamefont {J.~E.}\ \bibnamefont {Stalnaker}}, \bibinfo {author} {\bibfnamefont {N.~D.}\ \bibnamefont {Lemke}}, \bibinfo {author} {\bibfnamefont {N.}~\bibnamefont {Poli}}, \bibinfo {author} {\bibfnamefont {C.~W.}\ \bibnamefont {Oates}}, \bibinfo {author} {\bibfnamefont {T.~M.}\ \bibnamefont {Fortier}}, \bibinfo {author} {\bibfnamefont {S.~A.}\ \bibnamefont {Diddams}}, \bibinfo {author} {\bibfnamefont {L.}~\bibnamefont {Hollberg}}, \bibinfo {author} {\bibfnamefont {C.~W.}\ \bibnamefont {Hoyt}}, \bibinfo {author} {\bibfnamefont {A.~V.}\ \bibnamefont {Taichenachev}}, \ and\ \bibinfo {author} {\bibfnamefont {V.~I.}\ \bibnamefont {Yudin}},\ }\href {\doibase 10.1103/PhysRevLett.100.103002} {\bibfield  {journal} {\bibinfo  {journal} {Phys. Rev. Lett.}\ }\textbf {\bibinfo {volume} {100}},\ \bibinfo {pages} {103002} (\bibinfo {year} {2008})}\BibitemShut {NoStop}%
\bibitem [{\citenamefont {Moses}\ \emph {et~al.}(2023)\citenamefont {Moses}, \citenamefont {Baldwin}, \citenamefont {Allman}, \citenamefont {Ancona}, \citenamefont {Ascarrunz}, \citenamefont {Barnes}, \citenamefont {Bartolotta}, \citenamefont {Bjork}, \citenamefont {Blanchard}, \citenamefont {Bohn}, \citenamefont {Bohnet}, \citenamefont {Brown}, \citenamefont {Burdick}, \citenamefont {Burton}, \citenamefont {Campbell}, \citenamefont {Campora}, \citenamefont {Carron}, \citenamefont {Chambers}, \citenamefont {Chan}, \citenamefont {Chen}, \citenamefont {Chernoguzov}, \citenamefont {Chertkov}, \citenamefont {Colina}, \citenamefont {Curtis}, \citenamefont {Daniel}, \citenamefont {DeCross}, \citenamefont {Deen}, \citenamefont {Delaney}, \citenamefont {Dreiling}, \citenamefont {Ertsgaard}, \citenamefont {Esposito}, \citenamefont {Estey}, \citenamefont {Fabrikant}, \citenamefont {Figgatt}, \citenamefont {Foltz}, \citenamefont {Foss-Feig}, \citenamefont {Francois}, \citenamefont {Gaebler}, \citenamefont {Gatterman},
  \citenamefont {Gilbreth}, \citenamefont {Giles}, \citenamefont {Glynn}, \citenamefont {Hall}, \citenamefont {Hankin}, \citenamefont {Hansen}, \citenamefont {Hayes}, \citenamefont {Higashi}, \citenamefont {Hoffman}, \citenamefont {Horning}, \citenamefont {Hout}, \citenamefont {Jacobs}, \citenamefont {Johansen}, \citenamefont {Jones}, \citenamefont {Karcz}, \citenamefont {Klein}, \citenamefont {Lauria}, \citenamefont {Lee}, \citenamefont {Liefer}, \citenamefont {Lu}, \citenamefont {Lucchetti}, \citenamefont {Lytle}, \citenamefont {Malm}, \citenamefont {Matheny}, \citenamefont {Mathewson}, \citenamefont {Mayer}, \citenamefont {Miller}, \citenamefont {Mills}, \citenamefont {Neyenhuis}, \citenamefont {Nugent}, \citenamefont {Olson}, \citenamefont {Parks}, \citenamefont {Price}, \citenamefont {Price}, \citenamefont {Pugh}, \citenamefont {Ransford}, \citenamefont {Reed}, \citenamefont {Roman}, \citenamefont {Rowe}, \citenamefont {Ryan-Anderson}, \citenamefont {Sanders}, \citenamefont {Sedlacek}, \citenamefont
  {Shevchuk}, \citenamefont {Siegfried}, \citenamefont {Skripka}, \citenamefont {Spaun}, \citenamefont {Sprenkle}, \citenamefont {Stutz}, \citenamefont {Swallows}, \citenamefont {Tobey}, \citenamefont {Tran}, \citenamefont {Tran}, \citenamefont {Vogt}, \citenamefont {Volin}, \citenamefont {Walker}, \citenamefont {Zolot},\ and\ \citenamefont {Pino}}]{moses_race-track_2023}%
  \BibitemOpen
  \bibfield  {author} {\bibinfo {author} {\bibfnamefont {S.}~\bibnamefont {Moses}}, \bibinfo {author} {\bibfnamefont {C.}~\bibnamefont {Baldwin}}, \bibinfo {author} {\bibfnamefont {M.}~\bibnamefont {Allman}}, \bibinfo {author} {\bibfnamefont {R.}~\bibnamefont {Ancona}}, \bibinfo {author} {\bibfnamefont {L.}~\bibnamefont {Ascarrunz}}, \bibinfo {author} {\bibfnamefont {C.}~\bibnamefont {Barnes}}, \bibinfo {author} {\bibfnamefont {J.}~\bibnamefont {Bartolotta}}, \bibinfo {author} {\bibfnamefont {B.}~\bibnamefont {Bjork}}, \bibinfo {author} {\bibfnamefont {P.}~\bibnamefont {Blanchard}}, \bibinfo {author} {\bibfnamefont {M.}~\bibnamefont {Bohn}}, \bibinfo {author} {\bibfnamefont {J.}~\bibnamefont {Bohnet}}, \bibinfo {author} {\bibfnamefont {N.}~\bibnamefont {Brown}}, \bibinfo {author} {\bibfnamefont {N.}~\bibnamefont {Burdick}}, \bibinfo {author} {\bibfnamefont {W.}~\bibnamefont {Burton}}, \bibinfo {author} {\bibfnamefont {S.}~\bibnamefont {Campbell}}, \bibinfo {author} {\bibfnamefont {J.}~\bibnamefont {Campora}},
  \bibinfo {author} {\bibfnamefont {C.}~\bibnamefont {Carron}}, \bibinfo {author} {\bibfnamefont {J.}~\bibnamefont {Chambers}}, \bibinfo {author} {\bibfnamefont {J.}~\bibnamefont {Chan}}, \bibinfo {author} {\bibfnamefont {Y.}~\bibnamefont {Chen}}, \bibinfo {author} {\bibfnamefont {A.}~\bibnamefont {Chernoguzov}}, \bibinfo {author} {\bibfnamefont {E.}~\bibnamefont {Chertkov}}, \bibinfo {author} {\bibfnamefont {J.}~\bibnamefont {Colina}}, \bibinfo {author} {\bibfnamefont {J.}~\bibnamefont {Curtis}}, \bibinfo {author} {\bibfnamefont {R.}~\bibnamefont {Daniel}}, \bibinfo {author} {\bibfnamefont {M.}~\bibnamefont {DeCross}}, \bibinfo {author} {\bibfnamefont {D.}~\bibnamefont {Deen}}, \bibinfo {author} {\bibfnamefont {C.}~\bibnamefont {Delaney}}, \bibinfo {author} {\bibfnamefont {J.}~\bibnamefont {Dreiling}}, \bibinfo {author} {\bibfnamefont {C.}~\bibnamefont {Ertsgaard}}, \bibinfo {author} {\bibfnamefont {J.}~\bibnamefont {Esposito}}, \bibinfo {author} {\bibfnamefont {B.}~\bibnamefont {Estey}}, \bibinfo {author}
  {\bibfnamefont {M.}~\bibnamefont {Fabrikant}}, \bibinfo {author} {\bibfnamefont {C.}~\bibnamefont {Figgatt}}, \bibinfo {author} {\bibfnamefont {C.}~\bibnamefont {Foltz}}, \bibinfo {author} {\bibfnamefont {M.}~\bibnamefont {Foss-Feig}}, \bibinfo {author} {\bibfnamefont {D.}~\bibnamefont {Francois}}, \bibinfo {author} {\bibfnamefont {J.}~\bibnamefont {Gaebler}}, \bibinfo {author} {\bibfnamefont {T.}~\bibnamefont {Gatterman}}, \bibinfo {author} {\bibfnamefont {C.}~\bibnamefont {Gilbreth}}, \bibinfo {author} {\bibfnamefont {J.}~\bibnamefont {Giles}}, \bibinfo {author} {\bibfnamefont {E.}~\bibnamefont {Glynn}}, \bibinfo {author} {\bibfnamefont {A.}~\bibnamefont {Hall}}, \bibinfo {author} {\bibfnamefont {A.}~\bibnamefont {Hankin}}, \bibinfo {author} {\bibfnamefont {A.}~\bibnamefont {Hansen}}, \bibinfo {author} {\bibfnamefont {D.}~\bibnamefont {Hayes}}, \bibinfo {author} {\bibfnamefont {B.}~\bibnamefont {Higashi}}, \bibinfo {author} {\bibfnamefont {I.}~\bibnamefont {Hoffman}}, \bibinfo {author} {\bibfnamefont
  {B.}~\bibnamefont {Horning}}, \bibinfo {author} {\bibfnamefont {J.}~\bibnamefont {Hout}}, \bibinfo {author} {\bibfnamefont {R.}~\bibnamefont {Jacobs}}, \bibinfo {author} {\bibfnamefont {J.}~\bibnamefont {Johansen}}, \bibinfo {author} {\bibfnamefont {L.}~\bibnamefont {Jones}}, \bibinfo {author} {\bibfnamefont {J.}~\bibnamefont {Karcz}}, \bibinfo {author} {\bibfnamefont {T.}~\bibnamefont {Klein}}, \bibinfo {author} {\bibfnamefont {P.}~\bibnamefont {Lauria}}, \bibinfo {author} {\bibfnamefont {P.}~\bibnamefont {Lee}}, \bibinfo {author} {\bibfnamefont {D.}~\bibnamefont {Liefer}}, \bibinfo {author} {\bibfnamefont {S.}~\bibnamefont {Lu}}, \bibinfo {author} {\bibfnamefont {D.}~\bibnamefont {Lucchetti}}, \bibinfo {author} {\bibfnamefont {C.}~\bibnamefont {Lytle}}, \bibinfo {author} {\bibfnamefont {A.}~\bibnamefont {Malm}}, \bibinfo {author} {\bibfnamefont {M.}~\bibnamefont {Matheny}}, \bibinfo {author} {\bibfnamefont {B.}~\bibnamefont {Mathewson}}, \bibinfo {author} {\bibfnamefont {K.}~\bibnamefont {Mayer}},
  \bibinfo {author} {\bibfnamefont {D.}~\bibnamefont {Miller}}, \bibinfo {author} {\bibfnamefont {M.}~\bibnamefont {Mills}}, \bibinfo {author} {\bibfnamefont {B.}~\bibnamefont {Neyenhuis}}, \bibinfo {author} {\bibfnamefont {L.}~\bibnamefont {Nugent}}, \bibinfo {author} {\bibfnamefont {S.}~\bibnamefont {Olson}}, \bibinfo {author} {\bibfnamefont {J.}~\bibnamefont {Parks}}, \bibinfo {author} {\bibfnamefont {G.}~\bibnamefont {Price}}, \bibinfo {author} {\bibfnamefont {Z.}~\bibnamefont {Price}}, \bibinfo {author} {\bibfnamefont {M.}~\bibnamefont {Pugh}}, \bibinfo {author} {\bibfnamefont {A.}~\bibnamefont {Ransford}}, \bibinfo {author} {\bibfnamefont {A.}~\bibnamefont {Reed}}, \bibinfo {author} {\bibfnamefont {C.}~\bibnamefont {Roman}}, \bibinfo {author} {\bibfnamefont {M.}~\bibnamefont {Rowe}}, \bibinfo {author} {\bibfnamefont {C.}~\bibnamefont {Ryan-Anderson}}, \bibinfo {author} {\bibfnamefont {S.}~\bibnamefont {Sanders}}, \bibinfo {author} {\bibfnamefont {J.}~\bibnamefont {Sedlacek}}, \bibinfo {author}
  {\bibfnamefont {P.}~\bibnamefont {Shevchuk}}, \bibinfo {author} {\bibfnamefont {P.}~\bibnamefont {Siegfried}}, \bibinfo {author} {\bibfnamefont {T.}~\bibnamefont {Skripka}}, \bibinfo {author} {\bibfnamefont {B.}~\bibnamefont {Spaun}}, \bibinfo {author} {\bibfnamefont {R.}~\bibnamefont {Sprenkle}}, \bibinfo {author} {\bibfnamefont {R.}~\bibnamefont {Stutz}}, \bibinfo {author} {\bibfnamefont {M.}~\bibnamefont {Swallows}}, \bibinfo {author} {\bibfnamefont {R.}~\bibnamefont {Tobey}}, \bibinfo {author} {\bibfnamefont {A.}~\bibnamefont {Tran}}, \bibinfo {author} {\bibfnamefont {T.}~\bibnamefont {Tran}}, \bibinfo {author} {\bibfnamefont {E.}~\bibnamefont {Vogt}}, \bibinfo {author} {\bibfnamefont {C.}~\bibnamefont {Volin}}, \bibinfo {author} {\bibfnamefont {J.}~\bibnamefont {Walker}}, \bibinfo {author} {\bibfnamefont {A.}~\bibnamefont {Zolot}}, \ and\ \bibinfo {author} {\bibfnamefont {J.}~\bibnamefont {Pino}},\ }\href {\doibase 10.1103/PhysRevX.13.041052} {\bibfield  {journal} {\bibinfo  {journal} {Physical Review
  X}\ }\textbf {\bibinfo {volume} {13}},\ \bibinfo {pages} {041052} (\bibinfo {year} {2023})}\BibitemShut {NoStop}%
\bibitem [{\citenamefont {Muniz}\ \emph {et~al.}(2025)\citenamefont {Muniz}, \citenamefont {Stone}, \citenamefont {Stack}, \citenamefont {Jaffe}, \citenamefont {Kindem}, \citenamefont {Wadleigh}, \citenamefont {Zalys-Geller}, \citenamefont {Zhang}, \citenamefont {Chen}, \citenamefont {Norcia}, \citenamefont {Epstein}, \citenamefont {Halperin}, \citenamefont {Hummel}, \citenamefont {Wilkason}, \citenamefont {Li}, \citenamefont {Barnes}, \citenamefont {Battaglino}, \citenamefont {Bohdanowicz}, \citenamefont {Booth}, \citenamefont {Brown}, \citenamefont {Brown}, \citenamefont {Cairncross}, \citenamefont {Cassella}, \citenamefont {Coxe}, \citenamefont {Crow}, \citenamefont {Feldkamp}, \citenamefont {Griger}, \citenamefont {Heinz}, \citenamefont {Jones}, \citenamefont {Kim}, \citenamefont {King}, \citenamefont {Kotru}, \citenamefont {Lauigan}, \citenamefont {Marjanovic}, \citenamefont {Megidish}, \citenamefont {Meredith}, \citenamefont {McDonald}, \citenamefont {Morshead}, \citenamefont {Narayanaswami},
  \citenamefont {Nishiguchi}, \citenamefont {Paule}, \citenamefont {Pawlak}, \citenamefont {Pudenz}, \citenamefont {P\'erez}, \citenamefont {Ryou}, \citenamefont {Simon}, \citenamefont {Smull}, \citenamefont {Urbanek}, \citenamefont {van~de Veerdonk}, \citenamefont {Vendeiro}, \citenamefont {Wu}, \citenamefont {Xie},\ and\ \citenamefont {Bloom}}]{Muniz_Universal_YBGates}%
  \BibitemOpen
  \bibfield  {author} {\bibinfo {author} {\bibfnamefont {J.~A.}\ \bibnamefont {Muniz}}, \bibinfo {author} {\bibfnamefont {M.}~\bibnamefont {Stone}}, \bibinfo {author} {\bibfnamefont {D.~T.}\ \bibnamefont {Stack}}, \bibinfo {author} {\bibfnamefont {M.}~\bibnamefont {Jaffe}}, \bibinfo {author} {\bibfnamefont {J.~M.}\ \bibnamefont {Kindem}}, \bibinfo {author} {\bibfnamefont {L.}~\bibnamefont {Wadleigh}}, \bibinfo {author} {\bibfnamefont {E.}~\bibnamefont {Zalys-Geller}}, \bibinfo {author} {\bibfnamefont {X.}~\bibnamefont {Zhang}}, \bibinfo {author} {\bibfnamefont {C.-A.}\ \bibnamefont {Chen}}, \bibinfo {author} {\bibfnamefont {M.~A.}\ \bibnamefont {Norcia}}, \bibinfo {author} {\bibfnamefont {J.}~\bibnamefont {Epstein}}, \bibinfo {author} {\bibfnamefont {E.}~\bibnamefont {Halperin}}, \bibinfo {author} {\bibfnamefont {F.}~\bibnamefont {Hummel}}, \bibinfo {author} {\bibfnamefont {T.}~\bibnamefont {Wilkason}}, \bibinfo {author} {\bibfnamefont {M.}~\bibnamefont {Li}}, \bibinfo {author} {\bibfnamefont
  {K.}~\bibnamefont {Barnes}}, \bibinfo {author} {\bibfnamefont {P.}~\bibnamefont {Battaglino}}, \bibinfo {author} {\bibfnamefont {T.~C.}\ \bibnamefont {Bohdanowicz}}, \bibinfo {author} {\bibfnamefont {G.}~\bibnamefont {Booth}}, \bibinfo {author} {\bibfnamefont {A.}~\bibnamefont {Brown}}, \bibinfo {author} {\bibfnamefont {M.~O.}\ \bibnamefont {Brown}}, \bibinfo {author} {\bibfnamefont {W.~B.}\ \bibnamefont {Cairncross}}, \bibinfo {author} {\bibfnamefont {K.}~\bibnamefont {Cassella}}, \bibinfo {author} {\bibfnamefont {R.}~\bibnamefont {Coxe}}, \bibinfo {author} {\bibfnamefont {D.}~\bibnamefont {Crow}}, \bibinfo {author} {\bibfnamefont {M.}~\bibnamefont {Feldkamp}}, \bibinfo {author} {\bibfnamefont {C.}~\bibnamefont {Griger}}, \bibinfo {author} {\bibfnamefont {A.}~\bibnamefont {Heinz}}, \bibinfo {author} {\bibfnamefont {A.~M.~W.}\ \bibnamefont {Jones}}, \bibinfo {author} {\bibfnamefont {H.}~\bibnamefont {Kim}}, \bibinfo {author} {\bibfnamefont {J.}~\bibnamefont {King}}, \bibinfo {author} {\bibfnamefont
  {K.}~\bibnamefont {Kotru}}, \bibinfo {author} {\bibfnamefont {J.}~\bibnamefont {Lauigan}}, \bibinfo {author} {\bibfnamefont {J.}~\bibnamefont {Marjanovic}}, \bibinfo {author} {\bibfnamefont {E.}~\bibnamefont {Megidish}}, \bibinfo {author} {\bibfnamefont {M.}~\bibnamefont {Meredith}}, \bibinfo {author} {\bibfnamefont {M.}~\bibnamefont {McDonald}}, \bibinfo {author} {\bibfnamefont {R.}~\bibnamefont {Morshead}}, \bibinfo {author} {\bibfnamefont {S.}~\bibnamefont {Narayanaswami}}, \bibinfo {author} {\bibfnamefont {C.}~\bibnamefont {Nishiguchi}}, \bibinfo {author} {\bibfnamefont {T.}~\bibnamefont {Paule}}, \bibinfo {author} {\bibfnamefont {K.~A.}\ \bibnamefont {Pawlak}}, \bibinfo {author} {\bibfnamefont {K.~L.}\ \bibnamefont {Pudenz}}, \bibinfo {author} {\bibfnamefont {D.~R.}\ \bibnamefont {P\'erez}}, \bibinfo {author} {\bibfnamefont {A.}~\bibnamefont {Ryou}}, \bibinfo {author} {\bibfnamefont {J.}~\bibnamefont {Simon}}, \bibinfo {author} {\bibfnamefont {A.}~\bibnamefont {Smull}}, \bibinfo {author} {\bibfnamefont
  {M.}~\bibnamefont {Urbanek}}, \bibinfo {author} {\bibfnamefont {R.~J.~M.}\ \bibnamefont {van~de Veerdonk}}, \bibinfo {author} {\bibfnamefont {Z.}~\bibnamefont {Vendeiro}}, \bibinfo {author} {\bibfnamefont {T.-Y.}\ \bibnamefont {Wu}}, \bibinfo {author} {\bibfnamefont {X.}~\bibnamefont {Xie}}, \ and\ \bibinfo {author} {\bibfnamefont {B.~J.}\ \bibnamefont {Bloom}},\ }\href {\doibase 10.1103/PRXQuantum.6.020334} {\bibfield  {journal} {\bibinfo  {journal} {PRX Quantum}\ }\textbf {\bibinfo {volume} {6}},\ \bibinfo {pages} {020334} (\bibinfo {year} {2025})}\BibitemShut {NoStop}%
\bibitem [{\citenamefont {Hoyt}\ \emph {et~al.}(2005)\citenamefont {Hoyt}, \citenamefont {Barber}, \citenamefont {Oates}, \citenamefont {Fortier}, \citenamefont {Diddams},\ and\ \citenamefont {Hollberg}}]{Hoyt_Yb_Clock}%
  \BibitemOpen
  \bibfield  {author} {\bibinfo {author} {\bibfnamefont {C.~W.}\ \bibnamefont {Hoyt}}, \bibinfo {author} {\bibfnamefont {Z.~W.}\ \bibnamefont {Barber}}, \bibinfo {author} {\bibfnamefont {C.~W.}\ \bibnamefont {Oates}}, \bibinfo {author} {\bibfnamefont {T.~M.}\ \bibnamefont {Fortier}}, \bibinfo {author} {\bibfnamefont {S.~A.}\ \bibnamefont {Diddams}}, \ and\ \bibinfo {author} {\bibfnamefont {L.}~\bibnamefont {Hollberg}},\ }\href {\doibase 10.1103/PhysRevLett.95.083003} {\bibfield  {journal} {\bibinfo  {journal} {Phys. Rev. Lett.}\ }\textbf {\bibinfo {volume} {95}},\ \bibinfo {pages} {083003} (\bibinfo {year} {2005})}\BibitemShut {NoStop}%
\bibitem [{\citenamefont {Huntemann}\ \emph {et~al.}(2016)\citenamefont {Huntemann}, \citenamefont {Sanner}, \citenamefont {Lipphardt}, \citenamefont {Tamm},\ and\ \citenamefont {Peik}}]{Huntemann_Yb+_clock}%
  \BibitemOpen
  \bibfield  {author} {\bibinfo {author} {\bibfnamefont {N.}~\bibnamefont {Huntemann}}, \bibinfo {author} {\bibfnamefont {C.}~\bibnamefont {Sanner}}, \bibinfo {author} {\bibfnamefont {B.}~\bibnamefont {Lipphardt}}, \bibinfo {author} {\bibfnamefont {C.}~\bibnamefont {Tamm}}, \ and\ \bibinfo {author} {\bibfnamefont {E.}~\bibnamefont {Peik}},\ }\href {\doibase 10.1103/PhysRevLett.116.063001} {\bibfield  {journal} {\bibinfo  {journal} {Phys. Rev. Lett.}\ }\textbf {\bibinfo {volume} {116}},\ \bibinfo {pages} {063001} (\bibinfo {year} {2016})}\BibitemShut {NoStop}%
\bibitem [{\citenamefont {Ono}\ \emph {et~al.}(2022)\citenamefont {Ono}, \citenamefont {Saito}, \citenamefont {Ishiyama}, \citenamefont {Higomoto}, \citenamefont {Takano}, \citenamefont {Takasu}, \citenamefont {Yamamoto}, \citenamefont {Tanaka},\ and\ \citenamefont {Takahashi}}]{Ono_Yb_Kingplot}%
  \BibitemOpen
  \bibfield  {author} {\bibinfo {author} {\bibfnamefont {K.}~\bibnamefont {Ono}}, \bibinfo {author} {\bibfnamefont {Y.}~\bibnamefont {Saito}}, \bibinfo {author} {\bibfnamefont {T.}~\bibnamefont {Ishiyama}}, \bibinfo {author} {\bibfnamefont {T.}~\bibnamefont {Higomoto}}, \bibinfo {author} {\bibfnamefont {T.}~\bibnamefont {Takano}}, \bibinfo {author} {\bibfnamefont {Y.}~\bibnamefont {Takasu}}, \bibinfo {author} {\bibfnamefont {Y.}~\bibnamefont {Yamamoto}}, \bibinfo {author} {\bibfnamefont {M.}~\bibnamefont {Tanaka}}, \ and\ \bibinfo {author} {\bibfnamefont {Y.}~\bibnamefont {Takahashi}},\ }\href {\doibase 10.1103/PhysRevX.12.021033} {\bibfield  {journal} {\bibinfo  {journal} {Phys. Rev. X}\ }\textbf {\bibinfo {volume} {12}},\ \bibinfo {pages} {021033} (\bibinfo {year} {2022})}\BibitemShut {NoStop}%
\bibitem [{\citenamefont {Sanner}\ \emph {et~al.}(2019)\citenamefont {Sanner}, \citenamefont {Huntemann}, \citenamefont {Lange}, \citenamefont {Tamm}, \citenamefont {Peik}, \citenamefont {Safronova}, \citenamefont {Porsev} \emph {et~al.}}]{sanner_CloclLor}%
  \BibitemOpen
  \bibfield  {author} {\bibinfo {author} {\bibfnamefont {C.}~\bibnamefont {Sanner}}, \bibinfo {author} {\bibfnamefont {N.}~\bibnamefont {Huntemann}}, \bibinfo {author} {\bibfnamefont {R.}~\bibnamefont {Lange}}, \bibinfo {author} {\bibfnamefont {C.}~\bibnamefont {Tamm}}, \bibinfo {author} {\bibfnamefont {E.}~\bibnamefont {Peik}}, \bibinfo {author} {\bibfnamefont {M.~S.}\ \bibnamefont {Safronova}}, \bibinfo {author} {\bibfnamefont {S.~G.}\ \bibnamefont {Porsev}},  \emph {et~al.},\ }\href {\doibase 10.1038/s41586-019-1014-9} {\bibfield  {journal} {\bibinfo  {journal} {Nature}\ }\textbf {\bibinfo {volume} {567}},\ \bibinfo {pages} {204} (\bibinfo {year} {2019})}\BibitemShut {NoStop}%
\bibitem [{\citenamefont {Dreissen}\ \emph {et~al.}(2022)\citenamefont {Dreissen}, \citenamefont {Yeh}, \citenamefont {F{\"u}rst}, \citenamefont {Grensemann}, \citenamefont {Mehlst{\"a}ubler} \emph {et~al.}}]{dreissen_lorentz}%
  \BibitemOpen
  \bibfield  {author} {\bibinfo {author} {\bibfnamefont {L.~S.}\ \bibnamefont {Dreissen}}, \bibinfo {author} {\bibfnamefont {C.-H.}\ \bibnamefont {Yeh}}, \bibinfo {author} {\bibfnamefont {H.~A.}\ \bibnamefont {F{\"u}rst}}, \bibinfo {author} {\bibfnamefont {K.~C.}\ \bibnamefont {Grensemann}}, \bibinfo {author} {\bibfnamefont {T.~E.}\ \bibnamefont {Mehlst{\"a}ubler}},  \emph {et~al.},\ }\href {\doibase 10.1038/s41467-022-34818-0} {\bibfield  {journal} {\bibinfo  {journal} {Nature Communications}\ }\textbf {\bibinfo {volume} {13}},\ \bibinfo {pages} {7314} (\bibinfo {year} {2022})}\BibitemShut {NoStop}%
\bibitem [{\citenamefont {Hur}\ \emph {et~al.}(2022)\citenamefont {Hur}, \citenamefont {Aude~Craik}, \citenamefont {Counts}, \citenamefont {Knyazev}, \citenamefont {Caldwell}, \citenamefont {Leung}, \citenamefont {Pandey}, \citenamefont {Berengut}, \citenamefont {Geddes}, \citenamefont {Nazarewicz}, \citenamefont {Reinhard}, \citenamefont {Kawasaki}, \citenamefont {Jeon}, \citenamefont {Jhe},\ and\ \citenamefont {Vuletić}}]{hur_evidence_2022}%
  \BibitemOpen
  \bibfield  {author} {\bibinfo {author} {\bibfnamefont {J.}~\bibnamefont {Hur}}, \bibinfo {author} {\bibfnamefont {D.~P.}\ \bibnamefont {Aude~Craik}}, \bibinfo {author} {\bibfnamefont {I.}~\bibnamefont {Counts}}, \bibinfo {author} {\bibfnamefont {E.}~\bibnamefont {Knyazev}}, \bibinfo {author} {\bibfnamefont {L.}~\bibnamefont {Caldwell}}, \bibinfo {author} {\bibfnamefont {C.}~\bibnamefont {Leung}}, \bibinfo {author} {\bibfnamefont {S.}~\bibnamefont {Pandey}}, \bibinfo {author} {\bibfnamefont {J.~C.}\ \bibnamefont {Berengut}}, \bibinfo {author} {\bibfnamefont {A.}~\bibnamefont {Geddes}}, \bibinfo {author} {\bibfnamefont {W.}~\bibnamefont {Nazarewicz}}, \bibinfo {author} {\bibfnamefont {P.-G.}\ \bibnamefont {Reinhard}}, \bibinfo {author} {\bibfnamefont {A.}~\bibnamefont {Kawasaki}}, \bibinfo {author} {\bibfnamefont {H.}~\bibnamefont {Jeon}}, \bibinfo {author} {\bibfnamefont {W.}~\bibnamefont {Jhe}}, \ and\ \bibinfo {author} {\bibfnamefont {V.}~\bibnamefont {Vuletić}},\ }\href {\doibase
  10.1103/PhysRevLett.128.163201} {\bibfield  {journal} {\bibinfo  {journal} {Physical Review Letters}\ }\textbf {\bibinfo {volume} {128}},\ \bibinfo {pages} {163201} (\bibinfo {year} {2022})}\BibitemShut {NoStop}%
\bibitem [{\citenamefont {Lange}\ \emph {et~al.}(2021)\citenamefont {Lange}, \citenamefont {Peshkov}, \citenamefont {Huntemann}, \citenamefont {Tamm}, \citenamefont {Surzhykov},\ and\ \citenamefont {Peik}}]{R_Lange_lifetime}%
  \BibitemOpen
  \bibfield  {author} {\bibinfo {author} {\bibfnamefont {R.}~\bibnamefont {Lange}}, \bibinfo {author} {\bibfnamefont {A.~A.}\ \bibnamefont {Peshkov}}, \bibinfo {author} {\bibfnamefont {N.}~\bibnamefont {Huntemann}}, \bibinfo {author} {\bibfnamefont {C.}~\bibnamefont {Tamm}}, \bibinfo {author} {\bibfnamefont {A.}~\bibnamefont {Surzhykov}}, \ and\ \bibinfo {author} {\bibfnamefont {E.}~\bibnamefont {Peik}},\ }\href {\doibase 10.1103/PhysRevLett.127.213001} {\bibfield  {journal} {\bibinfo  {journal} {Phys. Rev. Lett.}\ }\textbf {\bibinfo {volume} {127}},\ \bibinfo {pages} {213001} (\bibinfo {year} {2021})}\BibitemShut {NoStop}%
\bibitem [{\citenamefont {Roman}(2021)}]{roman_expanding_2021}%
  \BibitemOpen
  \bibfield  {author} {\bibinfo {author} {\bibfnamefont {C.}~\bibnamefont {Roman}},\ }\emph {\bibinfo {title} {Expanding the $^{171}$Yb$^+$ toolbox: the $^2$F$_{7/2}$ state as resource for quantum information science}},\ \href@noop {} {Ph.D. thesis},\ \bibinfo  {school} {UCLA} (\bibinfo {year} {2021})\BibitemShut {NoStop}%
\bibitem [{\citenamefont {Yang}\ \emph {et~al.}(2022)\citenamefont {Yang}, \citenamefont {Ma}, \citenamefont {Wu}, \citenamefont {Wang}, \citenamefont {Cao}, \citenamefont {Guo}, \citenamefont {Huang}, \citenamefont {Feng}, \citenamefont {Zhou},\ and\ \citenamefont {Duan}}]{yang_realizing_2022}%
  \BibitemOpen
  \bibfield  {author} {\bibinfo {author} {\bibfnamefont {H.-X.}\ \bibnamefont {Yang}}, \bibinfo {author} {\bibfnamefont {J.-Y.}\ \bibnamefont {Ma}}, \bibinfo {author} {\bibfnamefont {Y.-K.}\ \bibnamefont {Wu}}, \bibinfo {author} {\bibfnamefont {Y.}~\bibnamefont {Wang}}, \bibinfo {author} {\bibfnamefont {M.-M.}\ \bibnamefont {Cao}}, \bibinfo {author} {\bibfnamefont {W.-X.}\ \bibnamefont {Guo}}, \bibinfo {author} {\bibfnamefont {Y.-Y.}\ \bibnamefont {Huang}}, \bibinfo {author} {\bibfnamefont {L.}~\bibnamefont {Feng}}, \bibinfo {author} {\bibfnamefont {Z.-C.}\ \bibnamefont {Zhou}}, \ and\ \bibinfo {author} {\bibfnamefont {L.-M.}\ \bibnamefont {Duan}},\ }\href {\doibase 10.1038/s41567-022-01661-5} {\bibfield  {journal} {\bibinfo  {journal} {Nature Physics}\ }\textbf {\bibinfo {volume} {18}},\ \bibinfo {pages} {1058} (\bibinfo {year} {2022})}\BibitemShut {NoStop}%
\bibitem [{\citenamefont {Allcock}\ \emph {et~al.}(2021)\citenamefont {Allcock}, \citenamefont {Campbell}, \citenamefont {Chiaverini}, \citenamefont {Chuang}, \citenamefont {Hudson}, \citenamefont {Moore}, \citenamefont {Ransford}, \citenamefont {Roman}, \citenamefont {Sage},\ and\ \citenamefont {Wineland}}]{OMG}%
  \BibitemOpen
  \bibfield  {author} {\bibinfo {author} {\bibfnamefont {D.~T.~C.}\ \bibnamefont {Allcock}}, \bibinfo {author} {\bibfnamefont {W.~C.}\ \bibnamefont {Campbell}}, \bibinfo {author} {\bibfnamefont {J.}~\bibnamefont {Chiaverini}}, \bibinfo {author} {\bibfnamefont {I.~L.}\ \bibnamefont {Chuang}}, \bibinfo {author} {\bibfnamefont {E.~R.}\ \bibnamefont {Hudson}}, \bibinfo {author} {\bibfnamefont {I.~D.}\ \bibnamefont {Moore}}, \bibinfo {author} {\bibfnamefont {A.}~\bibnamefont {Ransford}}, \bibinfo {author} {\bibfnamefont {C.}~\bibnamefont {Roman}}, \bibinfo {author} {\bibfnamefont {J.~M.}\ \bibnamefont {Sage}}, \ and\ \bibinfo {author} {\bibfnamefont {D.~J.}\ \bibnamefont {Wineland}},\ }\href@noop {} {\bibfield  {journal} {\bibinfo  {journal} {Applied Physics Letters}\ }\textbf {\bibinfo {volume} {119}},\ \bibinfo {pages} {214002} (\bibinfo {year} {2021})}\BibitemShut {NoStop}%
\bibitem [{\citenamefont {Tan}\ \emph {et~al.}(2021)\citenamefont {Tan}, \citenamefont {Edmunds}, \citenamefont {Milne}, \citenamefont {Biercuk},\ and\ \citenamefont {Hempel}}]{tan_precision_2021}%
  \BibitemOpen
  \bibfield  {author} {\bibinfo {author} {\bibfnamefont {T.~R.}\ \bibnamefont {Tan}}, \bibinfo {author} {\bibfnamefont {C.~L.}\ \bibnamefont {Edmunds}}, \bibinfo {author} {\bibfnamefont {A.~R.}\ \bibnamefont {Milne}}, \bibinfo {author} {\bibfnamefont {M.~J.}\ \bibnamefont {Biercuk}}, \ and\ \bibinfo {author} {\bibfnamefont {C.}~\bibnamefont {Hempel}},\ }\href {\doibase 10.1103/PhysRevA.104.L010802} {\bibfield  {journal} {\bibinfo  {journal} {Physical Review A}\ }\textbf {\bibinfo {volume} {104}},\ \bibinfo {pages} {L010802} (\bibinfo {year} {2021})},\ \bibinfo {note} {publisher: American Physical Society}\BibitemShut {NoStop}%
\bibitem [{\citenamefont {Campbell}(2025)}]{CampbellMultipoles}%
  \BibitemOpen
  \bibfield  {author} {\bibinfo {author} {\bibfnamefont {W.~C.}\ \bibnamefont {Campbell}},\ }\href {https://arxiv.org/abs/2510.07451} {\enquote {\bibinfo {title} {Angular geometry of atomic multipole transitions},}\ } (\bibinfo {year} {2025}),\ \Eprint {http://arxiv.org/abs/2510.07451} {arXiv:2510.07451 [quant-ph]} \BibitemShut {NoStop}%
\bibitem [{\citenamefont {Hur}(2022)}]{Hur_thesis}%
  \BibitemOpen
  \bibfield  {author} {\bibinfo {author} {\bibfnamefont {J.}~\bibnamefont {Hur}},\ }\emph {\bibinfo {title} {Probing New Physics with Spectroscopy of Trapped Ions}},\ \href@noop {} {Ph.D. thesis},\ \bibinfo  {school} {Massachusetts Institute of Technology} (\bibinfo {year} {2022})\BibitemShut {NoStop}%
\bibitem [{\citenamefont {Kramida}\ \emph {et~al.}(2024)\citenamefont {Kramida}, \citenamefont {{Yu.~Ralchenko}}, \citenamefont {Reader},\ and\ \citenamefont {{and NIST ASD Team}}}]{NIST_ASD}%
  \BibitemOpen
  \bibfield  {author} {\bibinfo {author} {\bibfnamefont {A.}~\bibnamefont {Kramida}}, \bibinfo {author} {\bibnamefont {{Yu.~Ralchenko}}}, \bibinfo {author} {\bibfnamefont {J.}~\bibnamefont {Reader}}, \ and\ \bibinfo {author} {\bibnamefont {{and NIST ASD Team}}},\ }\href@noop {} {}\bibinfo {howpublished} {{NIST Atomic Spectra Database (ver. 5.12), [Online]. Available: {\tt{https://physics.nist.gov/asd}} [2025, September 8]. National Institute of Standards and Technology, Gaithersburg, MD.}} (\bibinfo {year} {2024})\BibitemShut {NoStop}%
\bibitem [{\citenamefont {Shao}\ \emph {et~al.}(2023)\citenamefont {Shao}, \citenamefont {Yue}, \citenamefont {Ma}, \citenamefont {Huang}, \citenamefont {Guan},\ and\ \citenamefont {Gao}}]{shao_precision_2023}%
  \BibitemOpen
  \bibfield  {author} {\bibinfo {author} {\bibfnamefont {H.}~\bibnamefont {Shao}}, \bibinfo {author} {\bibfnamefont {H.}~\bibnamefont {Yue}}, \bibinfo {author} {\bibfnamefont {Z.}~\bibnamefont {Ma}}, \bibinfo {author} {\bibfnamefont {Y.}~\bibnamefont {Huang}}, \bibinfo {author} {\bibfnamefont {H.}~\bibnamefont {Guan}}, \ and\ \bibinfo {author} {\bibfnamefont {K.}~\bibnamefont {Gao}},\ }\href {\doibase 10.1103/PhysRevResearch.5.023193} {\bibfield  {journal} {\bibinfo  {journal} {Physical Review Research}\ }\textbf {\bibinfo {volume} {5}},\ \bibinfo {pages} {023193} (\bibinfo {year} {2023})},\ \bibinfo {note} {publisher: American Physical Society}\BibitemShut {NoStop}%
\bibitem [{\citenamefont {Kondev}\ \emph {et~al.}(2021)\citenamefont {Kondev}, \citenamefont {Wang}, \citenamefont {Huang}, \citenamefont {Naimi},\ and\ \citenamefont {Audi}}]{Kondev_2021}%
  \BibitemOpen
  \bibfield  {author} {\bibinfo {author} {\bibfnamefont {F.}~\bibnamefont {Kondev}}, \bibinfo {author} {\bibfnamefont {M.}~\bibnamefont {Wang}}, \bibinfo {author} {\bibfnamefont {W.}~\bibnamefont {Huang}}, \bibinfo {author} {\bibfnamefont {S.}~\bibnamefont {Naimi}}, \ and\ \bibinfo {author} {\bibfnamefont {G.}~\bibnamefont {Audi}},\ }\href {\doibase 10.1088/1674-1137/abddae} {\bibfield  {journal} {\bibinfo  {journal} {Chinese Physics C}\ }\textbf {\bibinfo {volume} {45}},\ \bibinfo {pages} {030001} (\bibinfo {year} {2021})}\BibitemShut {NoStop}%
\bibitem [{\citenamefont {Hucul}\ \emph {et~al.}(2017)\citenamefont {Hucul}, \citenamefont {Christensen}, \citenamefont {Hudson},\ and\ \citenamefont {Campbell}}]{Hucul_spectroscopy_2017}%
  \BibitemOpen
  \bibfield  {author} {\bibinfo {author} {\bibfnamefont {D.}~\bibnamefont {Hucul}}, \bibinfo {author} {\bibfnamefont {J.~E.}\ \bibnamefont {Christensen}}, \bibinfo {author} {\bibfnamefont {E.~R.}\ \bibnamefont {Hudson}}, \ and\ \bibinfo {author} {\bibfnamefont {W.~C.}\ \bibnamefont {Campbell}},\ }\href {\doibase 10.1103/PhysRevLett.119.100501} {\bibfield  {journal} {\bibinfo  {journal} {Phys. Rev. Lett.}\ }\textbf {\bibinfo {volume} {119}},\ \bibinfo {pages} {100501} (\bibinfo {year} {2017})}\BibitemShut {NoStop}%
\bibitem [{\citenamefont {Fan}\ \emph {et~al.}(2023)\citenamefont {Fan}, \citenamefont {Ready}, \citenamefont {Li}, \citenamefont {Kofford}, \citenamefont {Kwapisz}, \citenamefont {Holliman}, \citenamefont {Ladabaum}, \citenamefont {Gaiser}, \citenamefont {Griswold},\ and\ \citenamefont {Jayich}}]{Fan_laser_2023}%
  \BibitemOpen
  \bibfield  {author} {\bibinfo {author} {\bibfnamefont {M.}~\bibnamefont {Fan}}, \bibinfo {author} {\bibfnamefont {R.~A.}\ \bibnamefont {Ready}}, \bibinfo {author} {\bibfnamefont {H.}~\bibnamefont {Li}}, \bibinfo {author} {\bibfnamefont {S.}~\bibnamefont {Kofford}}, \bibinfo {author} {\bibfnamefont {R.}~\bibnamefont {Kwapisz}}, \bibinfo {author} {\bibfnamefont {C.~A.}\ \bibnamefont {Holliman}}, \bibinfo {author} {\bibfnamefont {M.~S.}\ \bibnamefont {Ladabaum}}, \bibinfo {author} {\bibfnamefont {A.~N.}\ \bibnamefont {Gaiser}}, \bibinfo {author} {\bibfnamefont {J.~R.}\ \bibnamefont {Griswold}}, \ and\ \bibinfo {author} {\bibfnamefont {A.~M.}\ \bibnamefont {Jayich}},\ }\href {\doibase 10.1103/PhysRevResearch.5.043201} {\bibfield  {journal} {\bibinfo  {journal} {Phys. Rev. Res.}\ }\textbf {\bibinfo {volume} {5}},\ \bibinfo {pages} {043201} (\bibinfo {year} {2023})}\BibitemShut {NoStop}%
\end{thebibliography}%

\end{document}